\input{aipcheck}

\documentclass[
    ,final            
  ]
  {aipproc}

\layoutstyle{6x9}

\begin{document}

\title{Medium Modifications from $^4$He$(\vec e,e' \vec p)^3$H}


\author{S. Malace}{
  address={University of South Carolina, Columbia, South Carolina 29208, USA}
}

\author{M. Paolone}{
  address={University of South Carolina, Columbia, South Carolina 29208, USA}
}

\author{S. Strauch for the Jefferson Lab Hall A Collaboration}{
  address={University of South Carolina, Columbia, South Carolina 29208, USA}
}

\begin{abstract}
  Polarization transfer in quasi-elastic nucleon knockout is sensitive
  to the properties of the nucleon in the nuclear medium, including
  possible modification of the nucleon form factor and/or spinor.  In
  our recently completed experiment E03-104 at Jefferson Lab we
  measured the proton recoil polarization in the $^4$He($\vec
  e,e^\prime \vec p\,$)$^3$H reaction at a $Q^2$ of 0.8 (GeV/$c$)$^2$
  and 1.3 (GeV/$c$)$^2$ with unprecedented precision. These data
   complement earlier data between 0.4 and 2.6 (GeV/$c$)$^2$ from both
  Mainz and Jefferson Lab. The measured ratio of polarization-transfer
  coefficients differs from a fully relativistic calculation, favoring
  either the inclusion of a medium modification of the proton form
  factors predicted by a quark-meson coupling model or strong
  charge-exchange final-state interactions. The measured induced
  polarizations agree well with the fully relativistic calculation and
  indicate that these strong final-state interactions may not be
  applicable.
\end{abstract}

\maketitle


\section{Introduction}
Whether the nucleon changes its fundamental properties while embedded in nuclear medium has been a long-standing question in nuclear physics, attracting experimental and 
theoretical attention.
Quantum chromodynamics (QCD) is established as the theory of the strong nuclear force but the degrees of freedom observed in nature, hadrons and nuclei, are 
different from those appearing in the QCD Lagrangian, quarks and gluons. There are no calculations available for nuclei within the QCD framework. Nuclei are 
effectively and well described as clusters of protons and neutrons held together by a strong, long-range force mediated by meson exchange \cite{mosz60}. Distinguishing possible changes in the structure of nucleons embedded in a nucleus from more conventional many-body effects like 
meson-exchange currents (MEC), isobar configurations (IC) or final-state interactions (FSI) is only possible within the context of a model. Therefore, interpretation of an 
experimental signature as an indication of nuclear medium modifications of the form factors makes sense only if this results in a more economical description of the nuclear 
many-body system.  In this context,  a calculation by Lu
{\it et al.}~\cite{Lu98}, using a quark-meson coupling (QMC) model,
suggests a measurable deviation from the free-space electromagnetic
form factor over the four-momentum-transfer squared $Q^2$ range 0.0 $< Q^2 <$ 2.5
(GeV/$c$)$^2$. Similar measurable effects have recently been
calculated in a light-front-constituent quark model by Frank {\it et
  al.} \cite{Fr96}, a modified Skyrme model by Yakshiev {\it et
  al.}~\cite{Ya02}, a chiral quark-soliton model by Smith and Miller
\cite{Smith04}, and the Nambu-Jona-Lasinio model of Horikawa and Bentz
\cite{Horikawa05}.  The connection between the modifications induced
by the nuclear medium of the nucleon form factors and of the deep
inelastic structure functions is discussed by Liuti \cite{Liuti06} 
using the concept of generalized parton distributions (GPDs). Guzey {\it et al.} \cite{guzev} study incoherent deeply virtual Compton scattering (DVCS) on $^4$He in the
 $^4{\rm He}(e,e^{\prime}\gamma p)X$ reaction, which probes medium-modifications of the bound nucleon GPDs and elastic form factors.

Experimental constraints for possible proton medium modifications are available for both the electric form factor $G_{E}$ and the magnetic form factor $G_{M}$. For example, 
the analysis of inclusive quasi-elastic scattering data on $^{3}$He showed that the deviation of the cross section from scattering from free nucleons scales to a function 
of a single variable $y$, the longitudinal momentum distribution. The $y$-scaling property is very sensitive to a change of nucleon radius. The interpretation of the 
$y$-scaling regime suggests 
that the nucleon radius is changed by less than 3 \%, at least for $Q^{2}$ values larger than 1 (GeV/$c$)$^{2}$ \cite{sickysc}. 
These measurements are in particular sensitive to the magnetic form factor. The interpretation of the Coulomb sum-rule measurements, sensitive rather to the electric form factor, 
is controversial: on $^{4}$He, good 
agreement between theory and experiment is obtained when using free-nucleon form factors \cite{carlson} and similar results were obtained on $^{12}$C and $^{56}$Fe \cite{jourd}. 
However, a re-examination of the extraction of the longitudinal and transverse response functions on medium-weight and heavy nuclei ($^{40}$Ca, $^{48}$Ca, $^{56}$Fe, $^{197}$Au,
 $^{208}$Pb and $^{238}$U) in the framework of the Effective Momentum 
Approximation (EMA) showed the quenching of the Coulomb sum rule. This corresponds to a relative change of the proton charge radius of 13 $\pm$ 4 \% and 
it was interpreted as an indication for a change of the nucleon properties inside the nuclear medium \cite{morg-meziani}.

Polarization transfer experiments of the type $(\vec e,e' \vec p)$ have measured polarization-transfer ratios which for a proton target are directly proportional to 
the ratio of the electric and magnetic form factors of the proton. 
When such measurements are performed on a nuclear target, the polarization-transfer observables are sensitive to the form-factor ratio of the proton embedded in the 
nuclear medium.
 In the following, the results of $^4$He$(\vec e,e^\prime \vec p)^3$H experiments to study the proton knock-out process will be
discussed. 

\section{Experiments}

The existing measurements of the polarization transfer and the induced polarization in $^4$He$(\vec e,e' \vec p)^3$H are the result of
three experiments. The first measurement was performed at the Mainz microtron (MAMI) at a $Q^{2}$-value of 0.4 (GeV/$c$)$^{2}$ \cite{mainz4he}. The Jefferson Lab experiment E93-049 extended the MAMI measurements 
to higher $Q^{2}$: 0.5, 1.0, 1.6 and 2.6 (GeV/$c$)$^{2}$ \cite{strauch03}. Our recent experiment E03-104 improved the $Q^{2}$ coverage by adding two 
high precision measurements at $Q^{2}$ = 0.8 and 1.3 (GeV/$c$)$^{2}$ \cite{e03104}. In view of decreasing other reaction mechanisms effects like MEC and FSI, 
the data were taken in quasi-elastic kinematics at low missing momentum with symmetry about the three-momentum-transfer
direction. In these experiments, two high-resolution spectrometers were used to detect the scattered electron 
and the recoil proton in coincidence. In addition, the missing-mass technique was used to identify $^3$H in the final state. First hand, 
we measure the azimuthal angular distribution of recoil protons that scatter via spin-orbit nuclear interaction on the graphite analyzer 
in a focal-plane polarimeter (FPP). The recoil proton polarization observables are then extracted from the azimuthal angular 
distributions taking into account the spin precession of the proton in magnetic fields \cite{Punjabi05}.

\begin{figure}[tb!]
  \includegraphics[angle=90,width=\textwidth]{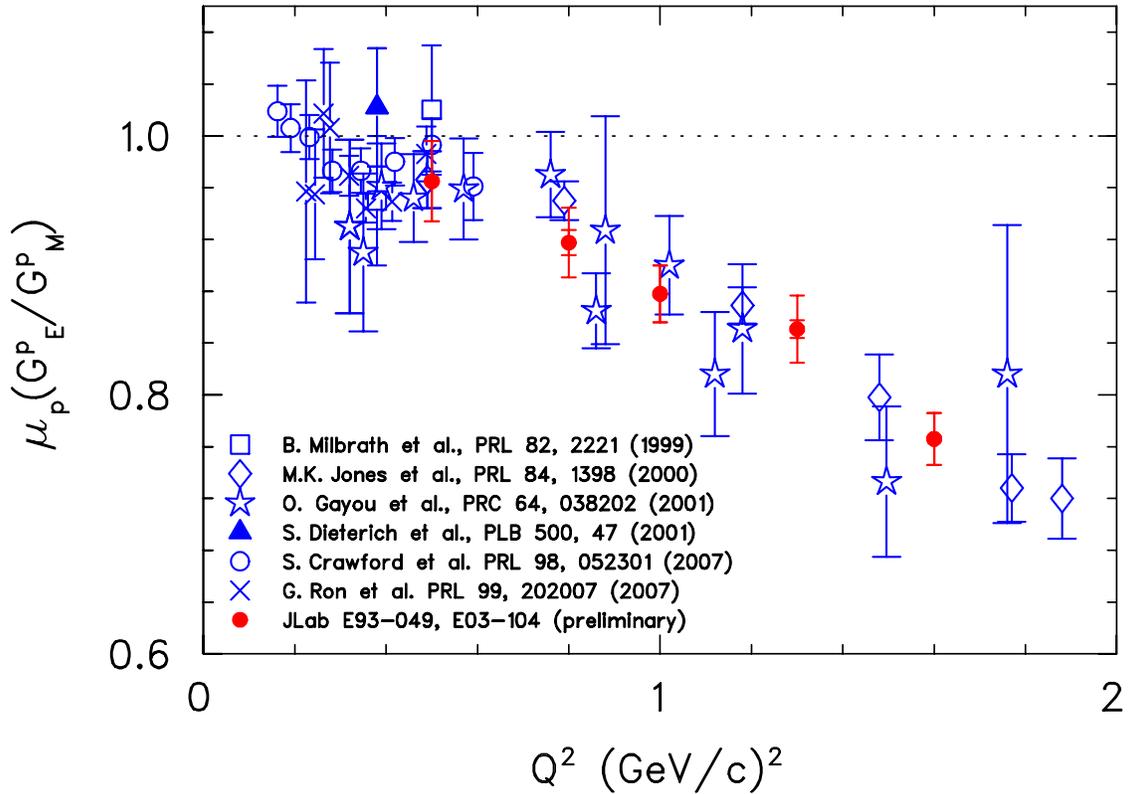}
  \caption{World data on the free proton form-factor ratio
    $G_E/G_M$ from recoil-polarization experiments as a function of $Q^2$. The Mainz (triangle) and
    Jefferson Lab experiment E93-049 points along with
    preliminary results from experiment E03-104 (filled circles)
    are in very good agreement with previous polarization-transfer data.
    For the preliminary data of E03-104 at $Q^{2}$ of 0.8 and 1.3 (GeV/$c$)$^{2}$, 
    the inner error bars are statistical while the outer error bars represent the total 
    preliminary uncertainties.
    \label{fig:gegm}}
\end{figure}

The $^{4}$He target was chosen because, in spite of its relative simplicity which allows for realistic microscopic theoretical 
calculations, it is rather dense, increasing possible medium modifications. In addition to the $^{4}$He target, a $^{1}$H
 target was also used in order to provide a base line for the comparison of in-medium to free proton polarizations. Therefore, our results on 
the proton polarization transfer are expressed in terms of the polarization-transfer double ratio, where the helium polarization ratio is 
normalized to the hydrogen polarization ratio, measured in an identical setting:

\begin{equation}
R = \frac{(P_x'/P_z')_{^4\rm He}}{(P_x'/P_z')_{^1\rm H}}.
\label{eq:rexp}
\end{equation}

As a by-product of the hydrogen measurements, the free proton form-factor ratio was extracted. Figure \ref{fig:gegm} shows our data to be in
 excellent agreement with previous polarization-transfer data \cite{Punjabi05,gep}. The E03-104 results at a $Q^{2}$ of 0.8 and 1.3 (GeV/$c$)$^{2}$ are still 
preliminary and in the final results we expect the systematic uncertainties to be greatly reduced in order to take full advantage of the impressive 
statistics available.

In the polarization-transfer double ratio $R$, nearly all systematic uncertainties cancel: the polarization-transfer observables are not 
 sensitive, to first order, to the instrumental asymmetries in the FPP, and their ratio is independent of the electron beam polarization and the 
graphite analyzing power. The small systematic uncertainties are due, mainly, to the uncertainties in the spin transport through the proton 
spectrometer but an extensive study is being performed in order to reduce their contribution to the total uncertainty.

The induced proton polarization $P_{y}$ is a direct measure of final-state interactions. However, the $P_{y}$ extraction is greatly 
complicated by the presence of instrumental asymmetries in the FPP. Typically, instrumental asymmetries are due to detector misalignments, 
detector inefficiencies or tracking algorithm issues. An ongoing effort aimed at devising a method to minimize instrumental asymmetries will 
make possible the precise extraction of the induced polarization $P_{y}$ from E03-104 measurements.   

\section{Results}

\begin{figure}[tb!]
  \includegraphics[angle=90,width=\textwidth]{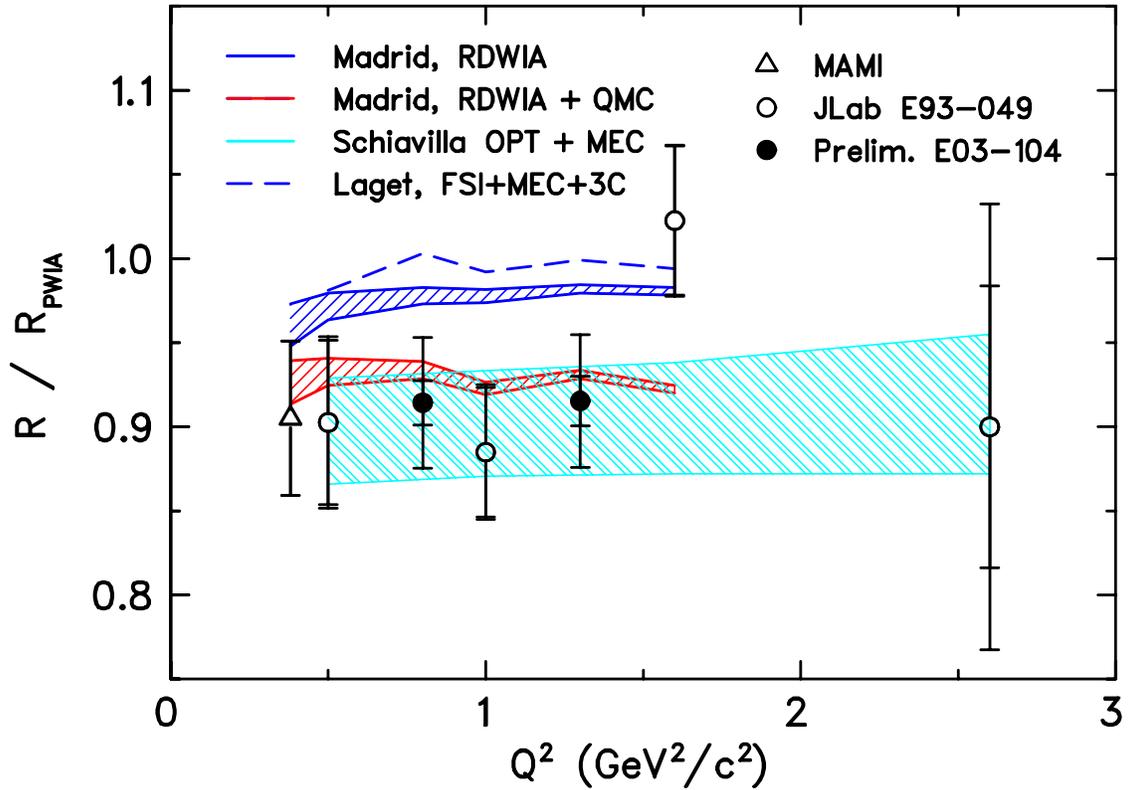}
  \caption{Superratio
    $R/R_{\rm PWIA}$ as a function of $Q^2$  from Mainz \cite{mainz4he} and
    Jefferson Lab experiment E93-049 \cite{strauch03} (open symbols) along with
    preliminary results from experiment E03-104 (filled circles); $R$
    is the ratio of transverse to longitudinal polarization of the
    recoiling proton in $^4$He$(\vec e,e^\prime \vec p)^3$H compared to the
    same ratio for $^1$H$(\vec e, e^\prime \vec p)$. The baseline
    $R_{\rm PWIA}$ is the value of $R$ obtained in a plane-wave
    calculation, to account for the 'trivial' effects of free {\it vs.} moving
    proton. The data are compared
    to calculations from the Madrid group \cite{Ud98}, Schiavilla
    {\it et al.} \cite{Schiavilla05} and Laget \cite{Laget94}. 
    \label{fig:ratio}}
\end{figure}

The polarization-transfer double ratio is shown in Figure \ref{fig:ratio}. The preliminary data from E03-104 (filled circles) are consistent with our
 previous data from E93-049 \cite{strauch03} and MAMI \cite{mainz4he} (open symbols). The inner error bars are statistical only. The final systematic uncertainties
 from E03-104 are expected to be much smaller than the ones shown here. The data are compared to several theoretical calculations. 

The Madrid fully relativistic distorted wave impulse approximation (RDWIA) \cite{Ud98} is a state of the art calculation which gives a fully relativistic treatment to
 the initial bound and final outgoing proton wave functions and to the current operator. The FSI are incorporated using relativistic optical potentials that distort the final nucleon 
wave function. MEC are not explicitly included in the Madrid calculation. Predictions by Meucci {\it et al.} \cite{meucci} show that the two-body current 
(the seagull diagram) effects are generally
small (less than 3 \% close to zero missing momenta) and visible only at high missing momenta. The blue band for Madrid RDWIA calculation (red band for RDWIA + QMC) 
is just a reflection of the theoretical uncertainties originating from different parameterizations of the current conservation prescriptions, cc1 and 
cc2 \cite{deforest}, or of the bound wave function. 
It can be seen that the Madrid RDWIA calculation (blue band) overpredicts the data by about 6 \%. After including the density-dependent medium-modified form factors 
as predicted by Lu {\it et al.} \cite{Lu98} in the RDWIA calculation (red band), good agreement with the data is obtained. 

This agreement has been interpreted as possible evidence of proton
medium modifications \cite{strauch03}. This interpretation is based on the description of the data in a particular 
model in terms of medium modifications of nucleon form factors and requires 
excellent control of the reaction mechanisms like meson-exchange currents, isobar configurations, and final-state interactions. 
In fact, there is an alternative interpretation of the observed suppression of the polarization-transfer ratio within a more traditional calculation 
by Schiavilla {\it et al.} \cite{Schiavilla05} (cyan band). Schiavilla's calculation uses free nucleon form factors and 
explicitly includes MEC effects which are suppressing $R$ by almost 4~\%. 
The FSIs are treated within the optical potentials framework and include both a spin-dependent and spin-independent charge-exchange term which play a 
crucial role in his calculation of P$_{y}$. The spin-independent charge-exchange term is 
constrained by $p$ + $^3$H $\to$ $n$ + $^3$He charge-exchange cross section data at kinetic lab energies of 57~MeV and 156~MeV. The spin-dependent 
charge-exchange term is taken to be real, with a depth parameter depending logarithmically on the kinetic lab energy, and with radius and diffuseness 
values of 1.2 fm and 0.15 fm, respectively.
In Schiavilla's model, the final-state interaction effects suppress $R$ by an additional 6 \% bringing this
calculation also in good agreement with the data.    
It should be stated that charge-exchange terms are not taken into account in the Madrid RDWIA calculation.

\begin{figure}
  \includegraphics[width=\textwidth]{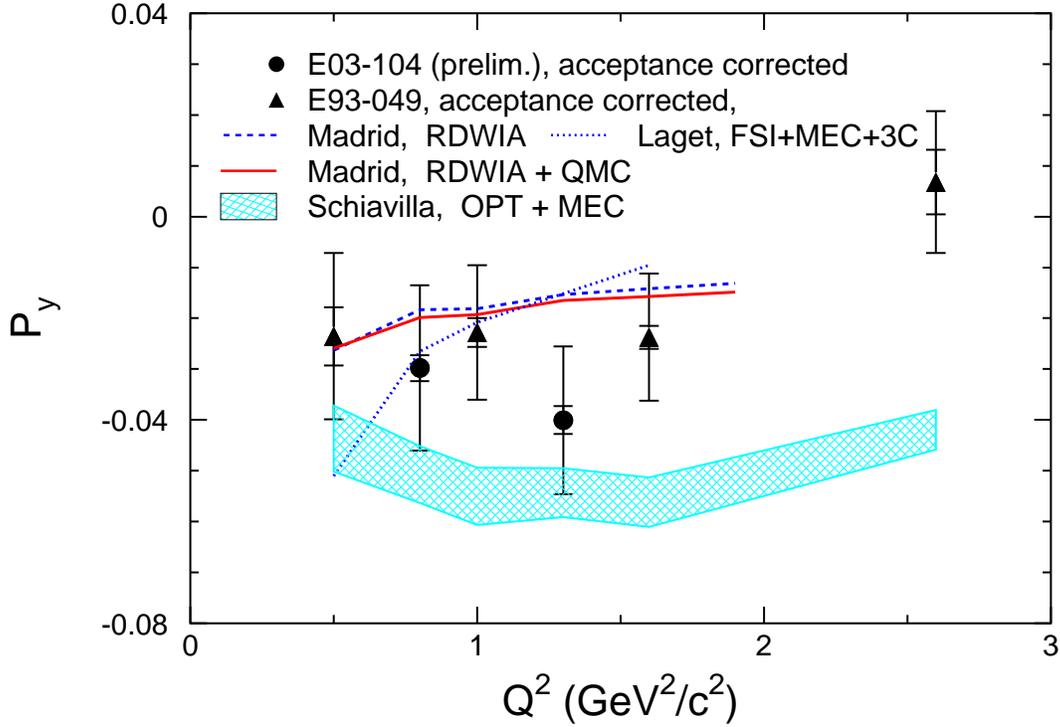}
  \caption{Induced polarization data from Mainz \cite{mainz4he} and
    Jefferson Lab experiment E93-049 \cite{strauch03} along with preliminary
    results from experiment E03-104. The data are compared to
    calculations from the Madrid group \cite{Ud98} and Schiavilla {\it et al.}
    \cite{Schiavilla05}. The comparison is made for missing momentum
    $p_m \approx 0$; note that the experimental data have been
    corrected for the spectrometer acceptance for this comparison.  }
  \label{fig:py}
\end{figure}

The difference in the modeling of final-state interactions is the
origin of the major part of the difference between the results of the
calculations by Madrid {\it et al.} \cite{Ud98} and Schiavilla {\it et
  al.}  \cite{Schiavilla05} for the polarization observables. Effects
from final-state interactions can be studied experimentally with the
induced polarization, $P_y$.  Figure \ref{fig:py} shows the data for
$P_y$. The induced polarization is small in this reaction. The sizable
systematic uncertainties are due to possible instrumental asymmetries.
Dedicated data have been taken during E03-104 to study these and we
hope to significantly reduce the systematic uncertainties in $P_y$ in
the final analysis. The data are compared with the results of the
calculations from the Madrid group and Schiavilla {\it et al.} at
missing momenta of about zero. The data have been corrected for the
spectrometer acceptance to facilitate this comparison.
The data suggest that the measured induced polarization (and thus the final-state interaction) is overestimated
in the model of Schiavilla {\it et al.}. Note that the charge-exchange terms, particularly, the 
spin-dependent one, gives the largest contribution to Schiavilla's calculation of P$_{y}$. 
The induced polarization proves to be sensitive to the choice of optical potential allowing this observable to be used to constrain theoretical models of FSI.

The results from Laget's calculation \cite{Laget94} are very similar to the PWIA ones. Laget's calculation overpredicts 
$R$ by almost 9 \% but is in fairly good agreement with the induced polarization measurements 
indicating a much stronger $Q^{2}$ dependence of FSI than the other models.  

A comparison of the model calculations in Figure \ref{fig:ratio} and Figure \ref{fig:py}
shows that the in-medium form factors (red band/curve) mostly affect the
ratio of polarization-transfer observables, not the induced
polarization. It is a great advantage of E03-104 to have access to both the polarization-transfer and the induced polarization.

In summary, polarization transfer in the quasi-elastic $(e,e^\prime p)$ reaction
is sensitive to possible medium modifications of the bound-nucleon
form factor, while at the same time largely insensitive to other
reaction mechanisms. Currently, the $^4$He$(\vec e, e^\prime \vec p)^3$H
polarization-transfer data can be well described by either the
inclusion of medium-modified form factors or strong charge-exchange
FSI in the models. However, these strong FSI effects may not be
consistent with the induced polarization data. The final analysis of
our new high-precision data from Jefferson Lab Hall A should provide a more
stringent test of these calculations.


\begin{theacknowledgments}
This work was supported in parts by the U.S. National Science Foundation: NSF PHY-0555604.
Jefferson Science Associates operates the Thomas Jefferson National Accelerator Facility under DOE
contract DE-AC05-06OR23177.
\end{theacknowledgments}

\end{document}